# Pleomorphic structural imperfections caused by pulsed Bi-implantation in the bulk and thin-film morphologies of TiO$_2$

D.A. Zatsepin[1,2], D.W. Boukhvalov[3,4], E.Z. Kurmaev[1,2], N.V. Gavrilov[5], S. S. Kim[6], I. S. Zhidkov[2]

[1]*M.N. Miheev Institute of Metal Physics of Ural Branch of Russian Academy of Sciences, 620990 Yekaterinburg, Russia*
[2]*Institute of Physics and Technology, Ural Federal University, 620002 Yekaterinburg, Russia*
[3]*Department of Chemistry, Hanyang University, 17 Haengdang-dong, Seongdong-gu, Seoul 04763, Republic of Korea*
[4]*Theoretical Physics and Applied Mathematics Department, Ural Federal University, Mira Street 19, 620002 Yekaterinburg, Russia*
[5]*Institute of Electrophysics, Russian Academy of Sciences, Ural Branch, 620990 Yekaterinburg, Russia*
[6]*School of Materials Science and Engineering, Inha University, Incheon 22212, Republic of Korea*

*The results of combined experimental and theoretical study of substitutional and clustering effects in Bi-doped TiO$_2$ hosts (bulk and thin-film morphologies) are presented. Bi-doping of the bulk and thin-film titanium dioxide was made with help of pulsed ion-implantation ($E_{Bi}^+ = 30$ keV, $D = 1 \times 10^{17}$ cm$^{-2}$) without posterior tempering. The X-ray photoelectron spectroscopy (XPS) qualification (core-levels and valence bands) and Density-Functional Theory (DFT) calculations were employed in order to study the electronic structure of Bi-ion implanted TiO$_2$ samples. According to XPS data obtained and DFT calculations, the Bi $\rightarrow$ Ti cation substitution occurs in Bi-implanted bulk TiO$_2$, whereas in the thin-film morphology of TiO$_2$:Bi the Bi-atoms have metal-like clusters segregation tendency. Based on the combined XPS and DFT considerations the possible reasons and mechanism for the observed effects are discussed. It is believed that established peculiarities of bismuth embedding into employed TiO$_2$ hosts are mostly the sequence of pleomorphic origin for the formed "bismuth–oxygen" chemical bonding.*

# 1. Introduction

A combination of bismuth and titanium oxides is rather prospective for various practical applications. Pure (undoped) $TiO_2$ usually is proposed as a technological base for a photoactive materials and bismuth oxide is rather attractive in the field of energy industry as a compound for solid oxide fuel cell cathodes (SOFC) and solid-state electrolytes [1-3]. Also it is used as a component of composite heterojunction photocatalysts [4-6] because of the relatively narrow band-gap ($E_g \sim 2.9$ eV), ionic type of conductivity (viz. mainly oxygen ions act as a charge carriers) and excellent photocatalytic performance under visible light excitation. Up to date four stable crystallographic polymorphs of "pure" $Bi_2O_3$ are known (α – δ) and available technologically as well as several metastable phases ($\epsilon$ and ω) [7-8]. The latter might be easily transformed into one of the stable oxide phases thru the variations of an ambient. Stable polymorphs of $Bi_2O_3$ demonstrate different crystal structures, ranging from monoclinic α-$Bi_2O_3$ up to the cubic fluorite-type δ-$Bi_2O_3$, but no changeover of Bi valence state had been established in these structural modifications – all the time Bi valence state remains 3+ [6-8]. With that, the conductivity character in some of these $Bi_2O_3$ phases is "switched" from *n*-type to *p*-type depending on the actual oxygen partial pressure (oxygen sublattice imperfections of vacant type) in the concrete polymorph, halfway impeding their employment as a durable material for energy industry.

From the foregoing follows, that the functional properties of bismuth oxide as an application material are strongly determined by the origin and type of Bi—O chemical bonding that is directly linked with the arrangement of oxygen atoms within the actual unit-cell from which the concrete polymorphic modification of $Bi_2O_3$ is assembled. At the same time well known, that in oxide materials the migration of oxygen and, hence, oxygen sublattice imperfections, are determined by the applied technology type ceteris paribus, on the contrary extending or limiting these phenomena. Thus it is believed that all the peculiarities mentioned will be possible to establish surely with the help of experimental techniques allowing probing directly the electronic structure of these interesting and prospective oxide materials (i.e. X-ray Photoelectron Spectroscopy, X-ray Emission

Spectroscopy, different variations of X-ray Absorption Spectroscopy, etc). In order to combine suitable $Bi_2O_3$ phases in terms of conductivity with the base compound of the newly fabricated photocatalyst (the final material is usually called Coupled Semiconductor – SC), the conventional technologies are applied, i.e. facile synthesis on the surface of oxide semiconducting host-matrix [4], ultrasonically dispersed deposition of previously synthesized $Bi_2O_3$ [6], hi-pressure tempering technology [9], material chemistry methods [10]. At least the following SC's were successfully synthesized with the help of reported techniques: $Bi_2O_3/NaBi(MoO_4)_2$, $\alpha$-$Bi_2O_3$/$\beta$-$Bi_2O_3$ and $Bi_2O_3/TiO_{2-x}N_x$ [4]. The latter seems to be very perspective, because a combination of Bi–O bonds inside the volume of $TiO_2$ (as a host-matrix) will strongly affect electronic band alignments of the initial $TiO_2$, allowing to elaborate the modification of charge-carrier transport properties of the newly developed heterojunction photocatalyst.

Well known, that ion-implantation is a powerful technique for ion-induced sintering of functional materials and electronic structure modification of the initial compound despite of some specific particular implantation cases, where the structure of modified host-matrix becomes imperfect and amorphous. Therefore a combination of an ion, selected for embedding, and target host-matrix structure have to be chosen in a strictly careful manner because of the possible critical mismatch of the ionic radii and posterior damage of initial host (so-called ballistic step of ion-implantation synthesis). Usually from the common considerations this step should be minimized in order to reduce the caused structure imperfections unless the appearance of luminescing defects, defects as precursors of a new growing phase or partial amorphyzation of the sample are specially required. So these initially mismatched combinations of implant and host-matrix might be quite attractive in terms of newly obtained electronic or other functional properties of the final material.

Concretely, the Bi-implantation of $TiO_2$ matrix seems rather promising because of the cited above accumulated experimental and theoretical data, but in this case we will face with a dramatic mismatch of ionic radii – 0.605 Å for Ti atom and 1.031 Å for Bi atom, and, presumably, with implantation stimulated overall final structure re-arrangement. Thus the results of this type of embedding and its peculiarities need a detailed study, and, preferably, by means of direct experimental and theoretical methods, effectively dealing with electronic and atomic structural configurations (particularly, the study of core-levels, valence bands and the re-configured electronic densities of states – DOS). The latter is the key-point for understanding and further developing of the functional properties because of the high mobility and polarisability of anion sub-lattice with the $6s^2$ lone-pair electrons of $Bi^{3+}$ in Bi–O containing materials [11-13].

The main purpose of the performed research is to determine the "as is" structural transformations of Bi-ion implanted $TiO_2$-host produced by pulsed mode of Bi-embedding and to establish the predecessors for the observed phenomena. That's why the combined experimental x-ray photoelectron and theoretical DFT studies of electronic structure and chemical bonding in Bi-doped $TiO_2$ in the bulk and thin-film morphologies without tempering were made and are presented in the current paper.

## 2. Experimental and computational details

Initial titanium dioxide host-matrices intended for Bi-ion implantation were synthesized using electrical explosion [14] and standard sol-gel chemical techniques in the form of bulk ceramics and thin-films, respectively. The employed for Bi-implantation the $TiO_2$-hosts were rutile single phase with tetragonal lattice parameters $a = 4.592$ Å and $c = 2.960$ Å and average crystallite size less than 200 nm. Sample dimensions were 13 mm in diameter 2 mm in height. The details of synthesis and posterior XRD-characterization of $TiO_2$ hosts are described in details in Refs. [15-16]. Bi-ion implantation of the bulk and thin-film $TiO_2$ samples was carried out in a pulsed-repetitive mode at the residual gas pressure of $1 \times 10^{-3}$ Pa with MEVVA-type ion source (see, for example Ref. [16] or other our previous papers). Bi-ion fluence (integrated flux) of $1 \times 10^{17}$ cm$^{-2}$ was achieved after 70 min of ion exposure. Bi-ion beam with ion energy of 30 keV, pulsed current density of not more than 0.8 mA/cm$^2$ and repetition rate of 12.5 Hz with pulse duration of 0.3 ms were applied for Bi-embedding. The average temperature of the samples during ion implantation did not exceed 295 °C in this case. The chemical purity of the initial and Bi-implanted $TiO_2$ samples was additionally qualified with the help of PHI XPS Versaprobe 500 spectrometer (ULVAC–Physical Electronics, USA), applying Al X-ray *Kα* radiation (1486.6 eV) [17]. The same XPS system had been employed for the core-level and valence band analysis of the samples under study. The results of XPS qualification (fast wide-scan or survey spectra) of Bi-implanted titanium dioxide host-matrices in the form of bulk and thin-film morphologies versus reference $TiO_2$ XPS external standard are presented at Fig.1.

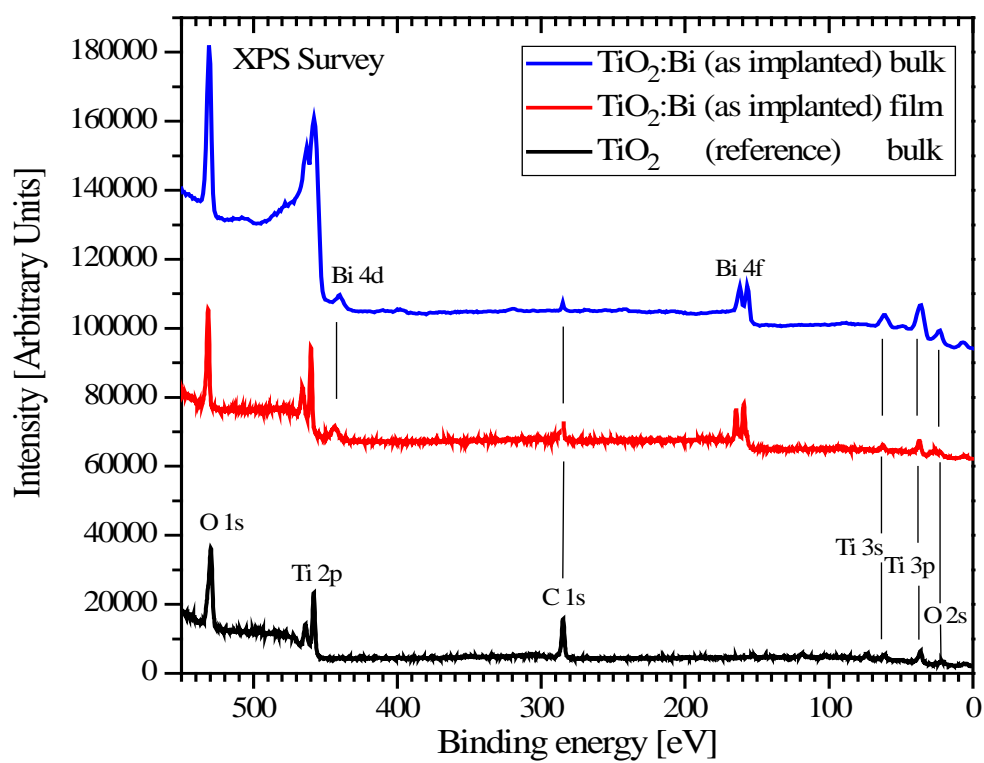

**Figure 1.** X-ray photoelectron (XPS) survey spectra of Bi-implanted titanium dioxide host-matrices in different morphologies and reference $TiO_2$ XPS external standard.

These spectra covers useful for chemical mapping the Binding Energy (BE) region, where majority of distinct and specific for the samples under study photoelectron peaks is arising. The detected and identified photoelectron peaks allow concluding that there are no alien impurities in implanted bulk and thin-film $TiO_2$ hosts except those belonging to embedded Bi-implant. Carbon 1*s* signal at 285.0 eV is relatively low comparing with the other XPS peaks but, at the same time, is well enough in order to perform precise calibration (see for details e.g. Ref.[17]) . Thereby we might conclude that there are no extra contaminations within the sensitivity range of applied XPS method and the declared empirical formulas for the samples under study are valid. The results of conventional estimation of average concentrations from our XPS spectra are presented in Table I.

**Table I.** Surface composition of Bi-ion implanted $TiO_2$ thin-film and ceramics (at.%).

| Sample under study | Ti (at.%) | O (at.%) | Bi (at.%) | C (at.%) |
|---|---|---|---|---|
| $TiO_2$:Bi (bulk morphology) | 29.7 | 55.3 | 5.9 | 9.2 |
| $TiO_2$:Bi (thin-film morphology) | 31.9 | 49.2 | 6.2 | 12.6 |

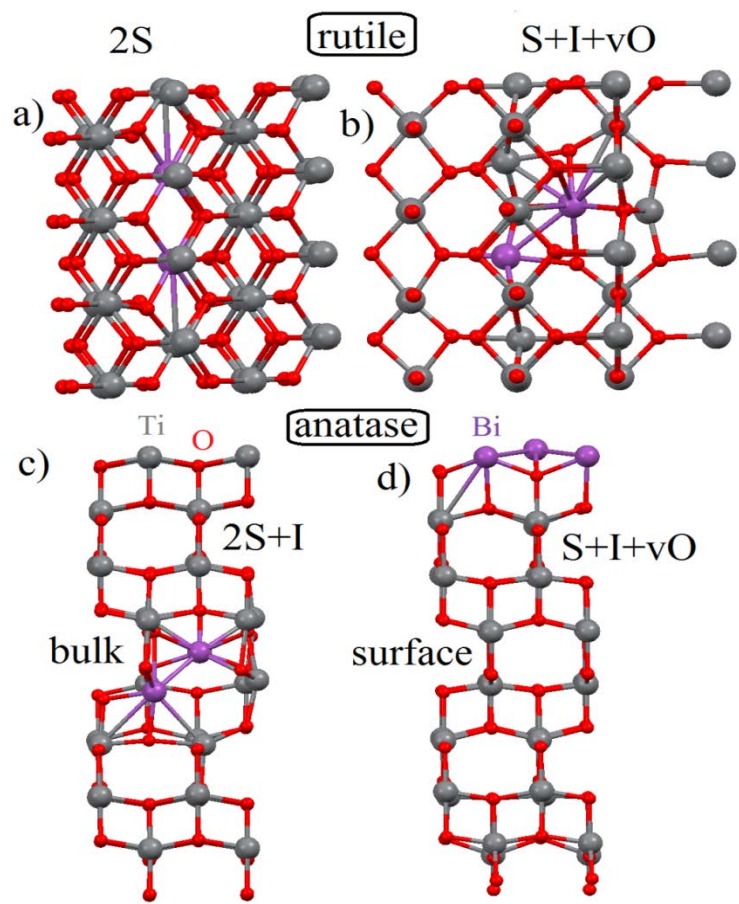

**Figure 2.** An optimized atomic structure of various configurations of substitutional (S) and interstitial (I) Bi-defects and oxygen vacancies (vO) in different $TiO_2$ hosts in bulk (a-c) and "surface" (d) - thin-film.

The valence density of states BE region (together with core-like bands), ranging from 0 eV up to ~ 30 eV, will be analyzed onwards in the current paper by means of XPS Valence Band Mapping approach.

The density-functional theory (DFT) calculations were performed using the SIESTA pseudopotential code [18-19] as had been previously utilized with a success for related studies of impurities in the bulk and thin-film semiconductors [17]. All calculations were made using the Perdew-Burke-Ernzerhof variant of the generalized gradient approximation (GGA-PBE) [19] for the exchange-correlation potential. A full optimization of the atomic positions was done, during which the electronic ground state was consistently found using norm-conserving pseudopotentials for the cores and a double-ξ plus polarization basis of localized orbitals for Bi, Ti, and O. The forces and total energies were optimized with an accuracy of 0.04 eV/Å and 1.0 meV, respectively. The calculations of the formation energies ($E_{form}$) were performed by considering the supercell both with and without a given defect [17]. The supercells consisting of 96 atoms were used as a host for studying defects in $TiO_2$ (see Fig. 2).

Taking into account our previous modelling of transition metal impurities in semiconductors [17], we have calculated various combinations of structural defects including single substitutional (1S) Bi-impurity, pairs of substitutional impurities (2S, Fig. 2a) and their combinations with interstitial (I) impurities (S + I and 2S + I) and oxygen vacancies (+vO, see Fig. 2b,c). For extended study of the case of bulk $TiO_2$ implantation we used the major structural phases of titanium dioxide – rutile and anatase. For the modeling of anatase-$TiO_2$ surface we applied the same supercell as a slab (Fig. 2d) which is feasible model of (001) surface of titanium dioxide [17].

## 3. Results and discussion

It seems logical to establish the formal valence state of Bi-implant after embedding into the bulk and thin-film $TiO_2$ host-matrices so XPS Bi 4*f* core-level spectra were recorded and analyzed (see Fig. 3). The Bi 4$f$ spectrum for α-Bi$_2$O$_3$ XPS external standard exhibits nearly symmetrical shapes of Bi 4$f_{7/2}$ – 4$f_{5/2}$ peaks, the BE value of 158.81 eV for Bi 4$f_{7/2}$ with spin-orbital 4$f_{7/2}$ – 4$f_{5/2}$ separation Δ = 5.29 eV and this is well agrees with the XPS data exactly for triangle α-Bi$_2$O$_3$ (158.78 eV and Δ ~ 5.2 eV, independently was reported in Refs. [6, 20] where both pure phase bismuth oxide polymorphs and mixed Bi$_2$O$_3$ phases had been studied). Thereof we might conclude about the high-quality of α-Bi$_2$O$_3$ XPS external standard employed for analysis. From Figure 3 it is distinctly seen that there is a good coincidence between Bi 4$f$ core-level spectra of TiO$_2$:Bi in the bulk morphology and α-Bi$_2$O$_3$, allowing to derive that the formal valence state of Bi in TiO$_2$:Bi is 3+, i.e. as in α-Bi$_2$O$_3$. The appropriate core-level spectrum of thin-film TiO$_2$:Bi displays two additional asymmetrical sub-bands at 156.69 eV and 161.91 eV, respectively, and these bands are absent in Bi 4$f$ core-level spectrum of α-Bi$_2$O$_3$. Taking into account their BE positions and the value of spin-orbital separation, these bands might indicate the allocation of metallic Bi-particles in thin-film TiO$_2$:Bi as metal clusters probably due to incomplete Bi-embedding into TiO$_2$ host-structure (compare XPS Bi-metal 4$f$ spectrum with that for the sample under study, Fig. 3). As for the main high-intensity 4$f$ peaks, they are the same as in α-Bi$_2$O$_3$ XPS external standard so bismuth formal valence states in thin-film TiO$_2$:Bi sample might be determined as Bi$^0$ and Bi$^{3+}$. The BEs of Bi-metal 4$f$ core-level spectrum precisely agrees with that reported in Refs. [21-22]. The effect of Bi spices segregation and partial allocation in the form of Bi-metal in Bi-doped TiO$_2$ was also reported by Wu *et.al.* [23] and well agrees with our XPS data.

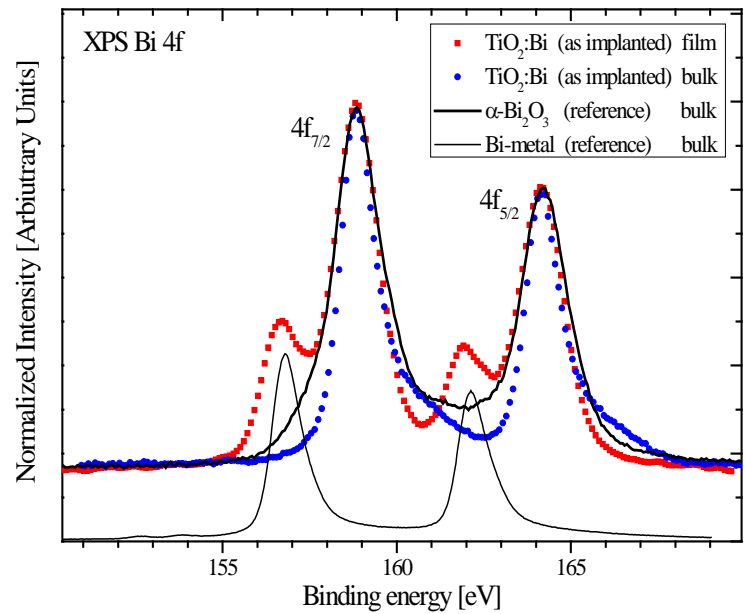

**Figure 3.** X-ray photoelectron (XPS) Bi 4*f* core-level spectra of Bi-implanted titanium dioxide host-matrices in different morphologies and references – Bi-metal and α-$Bi_2O_3$ XPS external standards.

According to Transmitting Electron Microscopy data on $Bi_2O_3$ phases (see e.g. [20]), an α-phase has essentially porous microstructure (average pore size of about 70 nm) versus even mixed α/β-phase, which is more compacted, so the most probable location of Bi-metal clusters might be inside these pores in our newly fabricated α-$Bi_2O_3$. Ghedia *et al.* also reporting about the large amount of empty space in crystalline structure of $Bi_2O_3$ phases depending on their sintering approach [9]. So the presence of pores might be considered as a possible point for limiting solid-state reactions of Bi-oxidation in our case. Note, that we can't declare about the formation of another $Bi_2O_3$-phase (i.e. β-$Bi_2O_3$, mixed α/β-$Bi_2O_3$ or others) exept "pure" alpha, because the rest bismuth oxides have strongly dissimilar spin-orbital separation in their XPS spectra [6,20,23] and this is supporting our XPS core-level analysis and conclusion about fabrication of α-$Bi_2O_3$ phase structural analogue both in the bulk and thin-film morphologies of $TiO_2$:Bi samples, but with mentioned above peculiarities.

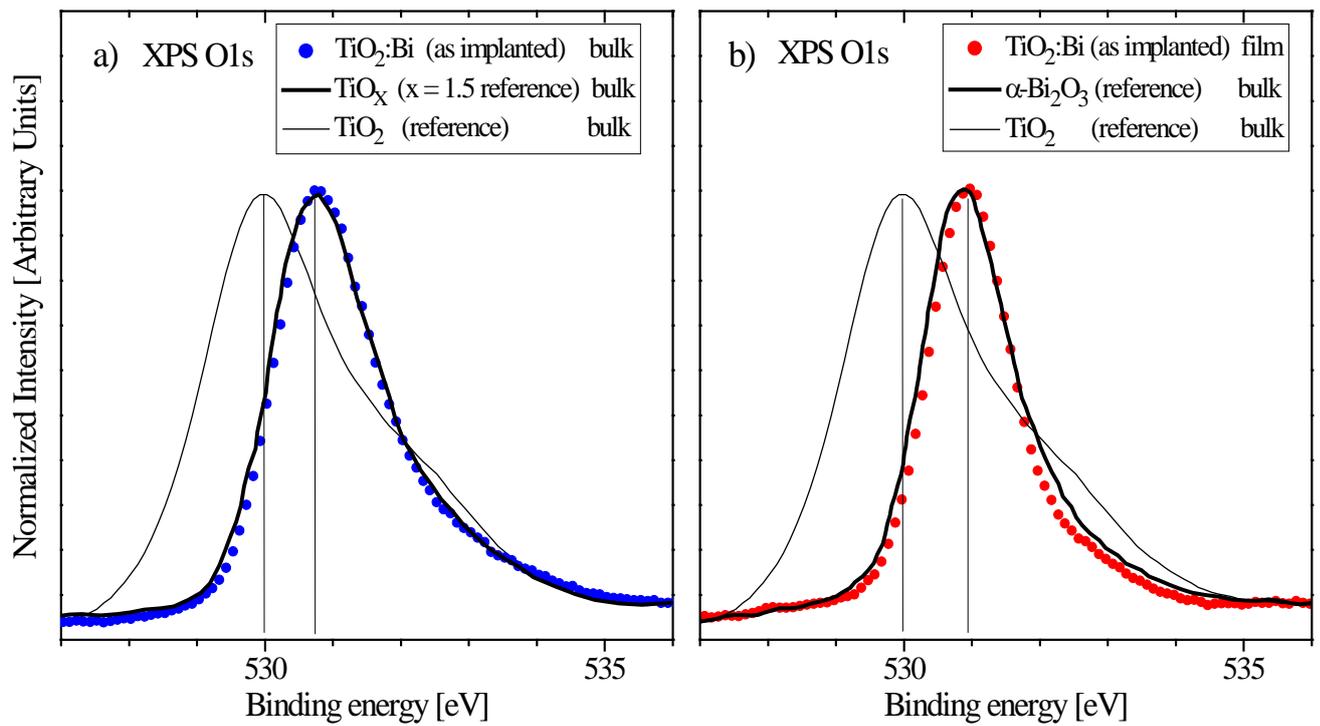

**Figure 4.** X-ray photoelectron (XPS) O 1*s* core-level spectra of Bi-implanted titanium dioxide host-matrices in different morphologies and references – $TiO_2$, $TiO_X$ (x = 1.5) and $\alpha$-$Bi_2O_3$ XPS external standards.

Formation of Bi–O bonds after embedding into titanium dioxide can occur because of the implantation stimulated distortion and disruption of initial Ti – O bonding with posterior re-arrangement of oxygen sublattice of Bi-implanted $TiO_2$-host. Usually these distortions of initial structure are well recognized by the XPS of appropriate core-levels because they appear as extrinsic reasons of variations in BE's value and full width at a half maximum (FWHM) of XPS spectrum [24]. Thus the XPS O 1*s* analysis will be useful for an identification of oxygen sublattice distortions caused by implantation (see Fig. 4). From Figure 4 the strong dissimilarities between O 1*s* XPS spectral parameters of initial $TiO_2$ and Bi-implanted $TiO_2$ (bulk and thin-film morphologies) are seen which, incidentally, are not surprising. The re-arranged oxygen sublattice means the re-arranged oxygen charge density surroundings of titanium and embedded bismuth atoms thus modifying the initial "metal–oxygen" hybridization [24-25]. As a result, the notable BE offset of ionized O 1s core-level occurs as well as the transformation of appropriate XPS spectrum shape. Based on reported above findings and taking into account the identity of XPS parameters for O 1s in $TiO_X$ reference and bulk $TiO_2$:Bi (Fig. 4a), we have a background to conclude about oxygen deficit emerged in Ti – O lattice by structural recombination of Ti and O atoms [26] following Bi-implantation and its posterior oxidation.

Another situation is arises in thin-film $TiO_2$:Bi (Fig. 4b), where the XPS O 1s core-level is identical to that for α-$Bi_2O_3$ and not for $TiO_X$ reference, because these spectra have notably different BE's values as well as FWHMs. At the same time the common feature of these spectra reveals in essential dissimilarity with XPS O 1s of initial $TiO_2$. We suppose that the reason for thin-film sample XPS O 1s identity with that for α-$Bi_2O_3$, but not for the bulk sample – $TiO_X$ reference, is concealed by exactly different mechanisms of Bi-embedding into different microstructure morphologies of $TiO_2$-host. In thin-film case we are observing the α-$Bi_2O_3$-like structure of O 1s and Bi 4f core-levels with Bi-metal contribution (Bi-loss effect), and the bulk case points out the defective $TiO_2$ structure where also an α-$Bi_2O_3$-phase is fabricating. Something-like close effect for Sn-implanted $TiO_2$ host-matrices in the bulk and thin-film morphologies was established and reported in Ref. [16]. Despite of the fact that in both analyzed cases there are the signs of an α-$Bi_2O_3$ phase fabrication, nonetheless there is a soundly reason to expect that the final microstructure might be different from what is reported here because of the initial dissimilar oxygen content (partial oxygen pressure) in thin-film and bulk, and, hence, from the very beginning, different conditions for Bi-embedding and posterior oxidation. This supposition is well supported by the existence of large number of $Bi_2O_3$ polymorphs with different atomic micro arrangement: several stoichiometric phases with various combinations of most common shortest (2.350 Å) and longest (2.635 Å) Bi – O bonding and low-symmetry oxygen vacant phases with distorted Bi – O bonding [9]. So visually different FWHM of XPS O 1s spectra for the bulk and thin-film $TiO_2$:Bi well agrees with our explanation because it is also the symptom of different from each other "metal-oxygen" interactions in thin oxide films and bulk oxides but of the same element composition [27].

We are supposing that the differences explored in O 1*s* core-level spectra will be also seen in O 2*s* and O 2*p* partial densities of states that will be highlighted by onward XPS Valence Band mapping.

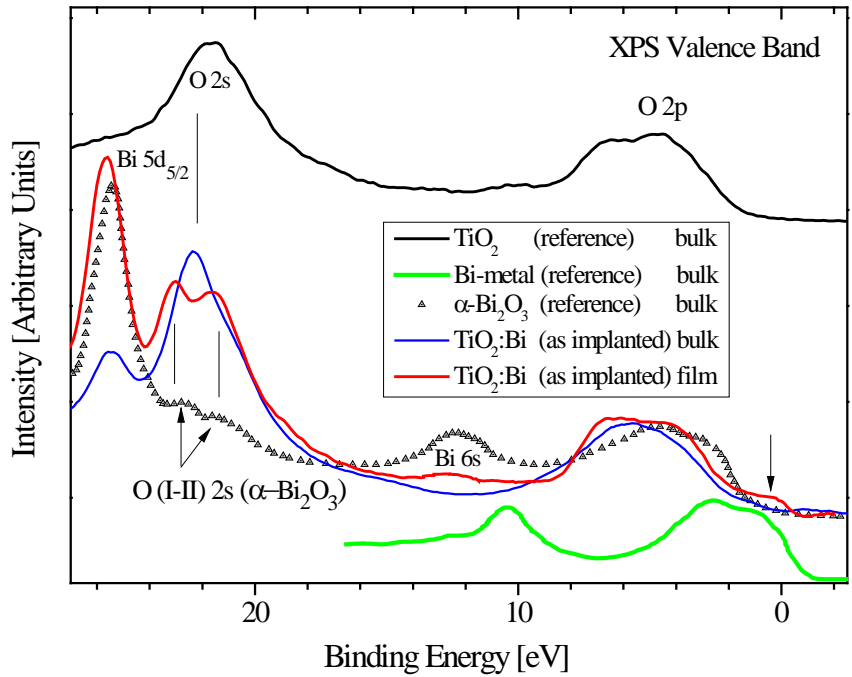

**Figure 5.** X-ray photoelectron (XPS) valence band spectra of Bi-implanted titanium dioxide host-matrices in different morphologies and references – Bi-metal, α-$Bi_2O_3$ and $TiO_2$ XPS external standards.

XPS Valence Band (VB) spectra of $TiO_2$:Bi samples in the form of bulk oxide and thin oxide film morphologies as well as Bi-metal, α-$Bi_2O_3$ and $TiO_2$ XPS external standards are presented for VB-mapping at Fig. 5. The most intensive XPS core-like band arising at ~ 25.5 eV in the VB-spectra of α-$Bi_2O_3$ and Bi-implanted $TiO_2$ host-matrices is absent in that for reference untreated $TiO_2$. This high-intensity XPS feature, having nearly identical XPS parameters [28] with our band, is also well-recognized spectrally in the valence band region of quasi-one-dimensional BiSi crystals. Additionally, according to NIST XPS Standard Reference Database [29], the ~ 25.5 eV band is present in appropriate spectra of all bismuth containing compounds, allowing to assign the origin of the recorded band with Bi $5d_{5/2}$ contribution to the VB electronic structure of our samples. The valence band BE region from about 21 eV up to 24 eV usually belongs to O 2*s* partial component of electronic structure for oxide compounds and, as it is seen from Fig. 5, indicates the strong transformation of XPS O 2$s$ band depending on the morphology of implanted TiO$_2$:Bi samples as well as when compared with the reference α-Bi$_2$O$_3$ and TiO$_2$. The monoclinic $P2_1/c$ α-Bi$_2$O$_3$ exhibiting an O 2$s$ contribution to the valence band as O 2$s$ (I) and O 2$s$ (II) XPS bands which are located respectively at around 22.8 eV and 21.4 eV. This finding does not contradict with the point of view that at least two strongly distinct types of Bi – O bonding usually are present in Bi$_2$O$_3$ polymorphs [9, 30] and with high-probability indicate the sequence of crystallographically independent oxygen sites have been detected by XPS VB mapping. At the same time the performed VASP calculations of partial oxygen charges from the converged electron density for Bi – O charge-transfer [31] display the sign for the third type of bismuth-oxygen bonding, but, as it was discussed in Refs. [9, 30], the actually calculated third-type charge-value for surrounding oxygen is rather close to one from the mentioned above distinct and crystallographically independent oxygen types (the so-called pleomorphic bonding). This third-type oxygen charge-value might occurs due to asymmetry imperfections in coordination of structural polyhedrons or/and light geometric distortions in α-Bi$_2$O$_3$. So it is possible that exactly for this reason we are observing only O 2$s$ (I) and O 2$s$ (II) XPS bands in VB spectrum of α-Bi$_2$O$_3$. In fact, the dissimilar contributions to the valence band structure of different O 2$s$ states for the case of Sn-implanted ZnO and TiO$_2$ host-matrices were reported in our previous XPS-study [16].

From Figure 5 it is seen that XPS VB spectrum for thin-film TiO$_2$:Bi is also exhibiting O 2$s$ (I) and O 2$s$ (II) states that are well coinciding with BE values for that in α-Bi$_2$O$_3$ XPS external standard. The Bi 6$s$ partial VB contributions are likewise for these compounds. With that, the observed higher intensity for O 2$s$ (I) and O 2$s$ (II) in XPS VB of thin-film TiO$_2$:Bi than in pure α-Bi$_2$O$_3$ phase might be due to Bi-metal loss effect in implanted thin oxide film, and this effect is supported by the good agreement of $E_F$ vicinity low-intensity and relatively wide shoulder at ~ 0.7 eV, which well agrees with the $E_F$ vicinity part of XPS VB spectrum of Bi-metal XPS external standard (see Fig. 5). Recall, that the same conclusion was made while analyzing the XPS Bi 4$f$ core-level spectra so there are no contradictions in discussed above data. The BE area in the valence band structure from about 2 eV up to 7.5 eV is believed to be contributed by the majority of O 2$p$ partial density of states and it is also different in shape from the same part of spectrum for pure α-Bi$_2$O$_3$, possibly indicating the influence of Bi-metal clustering and thus dissimilar arrangement of oxygen atoms in TiO$_2$ thin-film host after Bi-implantation. The very close situation with Bi-metal "droplets" in α-Bi$_2$O$_3$ is also reported in Ref.[32], where the direct SEM imaging of these droplets were presented as well as the XPS analysis of their final material. Comparing our technology approach of Bi-embedding into oxide host-matrix with self-catalyzed vapor-solid mechanism (SCVS) [17], we would like to note about the BiO$_X$-imperfections (oxygen deficient bismuth-oxygen clusters) in the final structure of Bi$_2$O$_3$ grown by SCVS. The authors of the paper [32] had detected these imperfections with the help of XPS O 1$s$ core-level analysis combined with SEM qualification of their Bi$_2$O$_3$ polymorph. From the strong dissimilarity of our XPS O 1$s$ core-level data for thin-film TiO$_2$:Bi and that after Ling *et. al.* [32], we might reasonably suppose that even if the BiO$_X$-imperfections are present in our thin-film TiO$_2$:Bi sample, they are not recognizable in our XPS data due to negligible concentration (at least within the well-known 0.2 at. % sensitivity limit of XPS method).

The situation with VB structure of bulk TiO$_2$:Bi is essentially different for thin-film morphology. Despite of the fact that XPS Bi 4$f$ core-level analysis points out the fabrication of pure α-Bi$_2$O$_3$ phase without Bi-metal clusters (see Fig. 3, no additional sub-bands at 156.69 eV and 161.91 eV), the TiO$_2$:Bi XPS VB spectrum is strongly dissimilar in the shape of main XPS features from that for α-Bi$_2$O$_3$ XPS external standard (Fig. 5). One can see the single O 2$s$ signal which rather close to that for reference TiO$_2$ than to O 2$s$ (I) or O 2$s$ (II) in α-Bi$_2$O$_3$ and thin-film TiO$_2$:Bi sample, also no ~ 0.7 eV $E_F$ vicinity shoulder exhibits from Bi-metal clustering and dramatically distorted O 2$p$ valence-band area. If we will take into account also the O 1$s$ analysis for the bulk case of TiO$_2$:Bi, then it is a reasonable background to suppose about the strongest overall structural re-arrangement of the initial TiO$_2$-host with the possible amorphyzation of the implanted bulk sample. In favor of this view-point the XPS detected oxygen deficient TiO$_X$ structural units are giving evidence (Fig. 4a) with relatively small-amount of α-$Bi_2O_3$ comparing with the concentration of $TiO_2$ and $TiO_X$ in the bulk morphology. In any way, the reported situation on Bi-implanted $TiO_2$ in the bulk morphology is not so easy for explanation only by means of XPS method, which is indicating the imperfections in the oxygen sub-lattice of initial $TiO_2$ and unlike anything VB transformation. The main differences in the fine structure of XPS VB of Bi-doped bulk and thin film $TiO_2$ are linked with additional peak appearance above the top of valence band for thin-film sample (indicated by arrow in the Fig. 5). For the explanation of this peak we have performed the model calculations of electronic structure of Bi-doped $TiO_2$.

In the first step of our calculations the evaluation of the energetics of formation the various configurations of Bi-impurities in the bulk and "surface" (thin-film) forms of $TiO_2$ hosts was made as well as the role of oxygen vacancies have been estimated. The results of the calculations performed are presented in Table II. The values obtained for both types of $TiO_2$ host are rather similar. It is seen from this table that oxygen vacancies play a significant role in the case of Bi-embedding in contrast with Sn-incorporation into the same host-matrices [16]. Note that in the case of denser rutile host the impact of vacancies is more valuable than in the case of anatase (note Fig. 2b,c). So far as Ti and Bi atoms have essentially different ionic radii – they differ almost nearly twice (0.605 and 1.03 Å, respectively), the appearance of Bi-impurity in the substitutional position is rather energetically unfavorable. At the same time the local violation of stoichiometry (i.e. oxygen vacancy, other type of interstitial or surface impurity) might significantly decrease the energy cost of Bi-insertion into $TiO_2$ matrix and, in this case, the inserted impurity can migrate from stoichiometric or nearly stoichiometric (pleomorphic) position that is crucial for an ion of large size. Also in the case of $TiO_2$:Bi the formed oxygen vacancies can play an additional role. Since bismuth could have only two oxidation states, namely 3+ and 5+, the direct substitution of $Ti^{4+}$ with $Bi^{3+}$ or $Bi^{5+}$ seems rather frustrating from this point of view, and oxygen deficit in the vicinity of paired substitutional Bi-impurities will "switch" the unnatural $Bi_2O_4$ configuration into ordinary $Bi_2O_3$ one. Additionally, Bi-impurity large radii effect might be considered as a preventing factor for metal-like clusters fabrication in the bulk morphology through the formation of corresponding bonds not only between the two Bi-impurities but also among interstitial impurities and oxygen ions in vicinity (see Fig. 2b). The calculated DOSes for selected from point of energetically favorable configurations (Fig. 6a) also demonstrate that despite of the band-gap decreasing, the electronic structure of $TiO_2$:Bi on the whole remains as $TiO_2$-like but with an additional Bi *5d* band at -24 eV. This theoretically derived result is in a good agreement with XPS VB spectra of bulk samples (Fig. 5).

**Table II.** Energies of formation (in eV per defect) of various types of defects including oxygen vacancies (vO) in vicinity. The most probable configurations for each type of host are marked with bold font.

| Defects configuration | Bulk Rutile (Anatase) | | "Surface" (Thin-film) | |
|---|---|---|---|---|
| | pure | +vO | pure | +vO |
| S | +8.03 (+7.89) | **+4.47 (+4.67)** | **-3.48** | +3.46 |
| 2S | +8.26 (+7.71) | +4.90 (+5.31) | -2.19 | +1.80 |
| S+I | +6.21 (+4.76) | **+4.67 (+4.66)** | -2.63 | **-0.08** |
| 2S+I | +6.02 (+5.17) | +5.01 (+5.09) | **-2.57** | **-0.11** |

Another dissimilar configuration of impurities could be realized for the surface (thin-films) morphology in contrast with the bulk one. This is quite important situation, especially for the case of "large" ions such as bismuth. Here the incorporation process of all types of Bi-impurities is energetically favorable (see Tab. II). Similarly to the case of bulk, the presence of oxygen vacancies provides significant changes in the energetics of defects, namely, an increasing of formation energies because the creating of some additional space on the surface for large ions is not so crucial comparing with the bulk. The calculated formation energies demonstrate that in this case the combination of substitutional and interstitial impurities (Fig. 1c) will be favorable in the form of clusters.

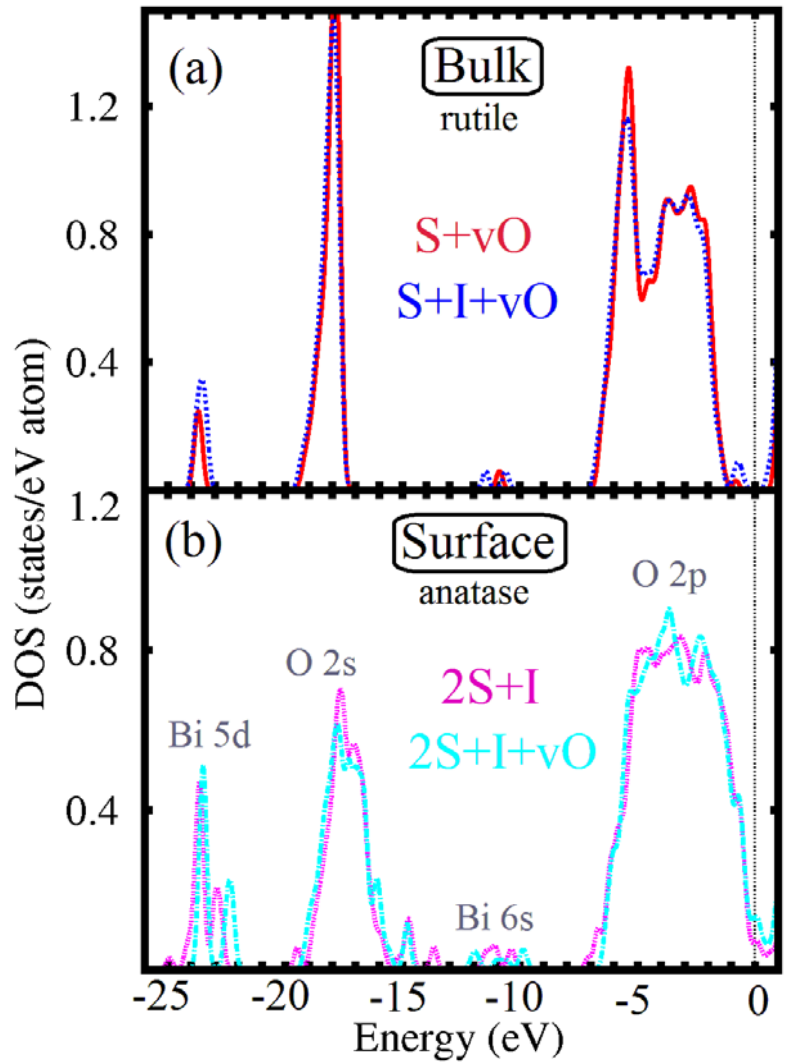

**Figure 6.** Total Densities of States (DOS) of the most probable configuration of defects (a) in the bulk and (b) in the "surface" (thin-film) morphologies of $TiO_2$ host.

With that, the electronic structure of these clusters (Fig. 6b) is in good agreement with XPS VB spectra of thin-films (Fig. 5). All three main XPS features in XPS VB of thin-films are clearly reproduced in the electronic structure of 2S+I(+vO) clusters on $TiO_2$ surface and they are: *i*. the appearance of DOSes on Fermi level, *ii*. some visible contribution at -10 eV and *iii*. the broadening of O 2*s* peak at -18 eV. Thus we might conclude that in contrast to bulk, where Bi-impurities form configurations with oxide-like atomic and electronic structure, in thin-films the segregation of Bi-impurities occurs in such way, that the clusters with metal-like atomic and electronic structures appear.

## 4. Conclusions

The Bi-doped TiO$_2$ hosts in the bulk and thin-film morphologies were studied with the use of XPS and DFT methods. It was established that both in the bulk and thin-film morphologies the α-Bi$_2$O$_3$-like phase is forming under Bi-ion pulsed implantation, but, as a dissimilarity, the Bi-metal loss was surely found in TiO$_2$:Bi thin-films. The formal valence states of Bi in our samples after implantation were detected as Bi$^0$ and Bi$^{3+}$. No signs of other valence states of Bi, which were the subject of some discussions (see e.g. Refs. [33-34]), had been detected in our combined study. It was shown, that the dissimilarities in Bi-implantation into TiO$_2$ hosts are mostly linked with a pleomorphic origin of "bismuth-oxygen" bonding, causing the formation of oxygen deficient bismuth-oxygen clusters as a precursors for Bi-loss and metal-like clusters segregation tendency in thin-films. In the bulk form of TiO$_2$ hosts the strongly defective TiO$_X$ structure was found which arises due to dramatical mismatch of ionic radii of Ti and Bi atoms. The results of DFT-calculations demonstrate the significant role of oxygen vacancies in the formation of various configurations of Bi-defects and crucial difference between bismuth incorporation into the bulk and "surface" forms. In the case of bulk morphology the formation of Bi$_2$O$_3$-like structure is more energetically favorable in contrast to the "surface" where the aggregation of metallic bismuth is essentially preferable.

**Acknowledgements**

This study was supported by the Act 211 of the Government of the Russian Federation (Agreement No. 02.A03.21.0006) and the Government Assignment of Russian Ministry of Education and Science (Contract No. 3.1016.2014/K).